# Optimal control of signalized intersection using hierarchical fuzzy-real control

M. Shahgholian[1], D. Gharavian[1*]

1: Department of Electrical and Computer Engineering, Shahid Beheshti University, Tehran, Iran
*D_Gharavian@sbu.ac.ir

**Abstract:** This paper presents a method based on precise modeling of traffic flow using fuzzy-real algorithm for optimal control of a signalized intersection. By improving social indicators such as security, well-fare and economy, intercity transportation has increased and traffic congestion developed. Today, the plight of traffic congestion is the cause of waste of time, capital, and health risks. Although construction of new streets, expressways, and highways can improve this situation, when inadequate space and limited budget is available, it is not practical. Optimization of signalized intersections for optimal use of the network capacity is proposed as a useful solution to resolve an important part of the problems developed at intersections. In this paper, with the aim of assessing the algorithms proposed for controlling signalized intersections and given the real potentials of common sensors present in traffic control systems, a suitable simulation medium called Signalized Intersection Fuzzy-Real Control (SIFRC) has been designed and implemented. Through the possibility made, the Abshar intersection in Isfahan city was modeled and phased through six methods: fixed time, pre-time control, segmental pre-time, fuzzy, real-time, and fuzzy-real. The results indicated better performance of the fuzzy-real method compared to the fixed time method up to 50%, and up to 10% compared to other methods. In addition, the speed of achieving the optimal response in the fuzzy-real method has been 11 times as large as that of the real-time method, while the delay time developed by the intersection in this method did not increase compared to the real-time method.

## 1. Introduction

The major difference between traffic and other social phenomena is its inverse growth. Increased population of cities and easier access to vehicles have caused the plight of traffic congestion to change into a major repetitive problem in metropolises of the world. Research has shown that vehicles, during congestion conditions, consume up to 80% more fuel compared to free conditions, thereby aggravating the air pollution[1]. Evidently, low air quality causes respiratory diseases such as asthma and bronchitis, and also increases the probability of incidence of dangerous diseases such as cancer. Annually, emission of particulate matters alone is the cause of around 30,000 premature deaths in the world[2].

On the other hand, according to a recent report by Inrix company, drivers in Los Angeles city, Moscow, London, and Paris waste around 102, 91, 74, and 69 h of their time respectively in traffic congestion every year[3]. On average, the cost of annual traffic congestions for Los Angeles, New York, London, and Berlin has been estimated to be 19.2, 33.7, 12.2, and 7.5 billion dollars respectively. Overall, in 2018, the economy of the US, Germany, and England incurred loss of around 416 billion dollars because of traffic congestion[4].

Increasing the capacity and developing the street network in cities concerning the densities of buildings and the volume of transportation cannot be done in many urban regions. On the other hand, implementation of this approach, given its staggering costs, requires precise recognition of the network and proper planning. So far, engineers have presented various solutions to improve the traffic situation in cities. It can be stated that all of these solutions have two objectives: 1. Shortening the travel time of the urban network users, 2. Optimal use of the current network capacity[5]. Accordingly, optimization of the current at-grade intersections is proposed as a suitable solution for resolving part of the problems developed in the traffic network[6]. The urban traffic network is a dynamic system with an uncertain nature and a nonlinear subsystem dependence along with a large number of variables including the rate of flow of vehicles, the length of the lines, and the scheduling of phases[7]. Given these complexities, usage of smart algorithms is essential to optimize the performance of signalized intersections. Note that traffic lights are installed at intersections where the sheer volume of traffic may prevent efficient and safe use of the intersection. In order to determine the performance of traffic lights, three measures can be taken: 1. Fixed time 2. Pre-time control 3. Traffic responsive.

The fixed time method and pre-time control use a predetermined cycle and phase. This method is suitable only for stable and regular traffic flows. Webster and Miller in 1958 and 1963 respectively presented a computational method and time model in order to minimize the delay time of vehicles. This method created a framework for modern traffic signal controls[8], [9]. With advances in sensors, real-time traffic responsive control was introduced. The main characteristic of this method is its real-time performance based on the current traffic information. In other words, in this method, controlling strategy is designed online for the traffic conditions of that moment. Trabia (1999) stated a two-stage fuzzy controller for a single intersection. He obtained the traffic data using loop sensors installed at specific intervals in the entrance of intersections. These sensors estimated the passing flow and the magnitude of the queue of each street. In the first stage, the data observed from the traffic flow were used in order to estimate the traffic intensity of two streets. Then, these traffic intensities determined whether the current phase magnitude should be longer or shorter[10]. In 2001, Bingham propounded a fuzzy controller with simple membership functions in order to control traffic lights. The membership functions of this controller were updated by reinforcement learning algorithm, where the functions offering good response were reinforced, while those generating a bad response were punished. This reinforcement causes these functions to be used in subsequent stages. Simulations indicated better performance of this method for fixed volumes of traffic[11]. In 2000 and 2004, Wiering presented an algorithm based on multifactorial reinforce learning for controlling traffic lights. Concerning the stop time estimated for vehicle by value functions, the reinforcement learning (RL) system considers a specific set for controlling traffic lights. In this algorithm, every traffic light is considered as a factor, and the relations between factors are used for selecting a suitable controlling set of lights. Further, these value functions were used to guide drivers in selecting the optimal path. The results of tests showed effective performance of the algorithm especially regarding correction of the selected path[12], [13]. Bazzan (2005) used an evolutionary game theory to develop a distributed outcome for coordinating traffic signal controller agents. Every agent plays a two-player game against each member of its neighbor, such that it finds Nash equilibrium under stable evolutionary strategy. There is no need for the agents to know the strategy of their opinions. Concerning the low capacity of information transmission of the network, the outcome of this method is more reliable[14]. In 2010, Alvarez and Pozniak used an uncooperative game outcome in order to optimize the timing of traffic lights. They considered every intersection as a non-cooperative game and stated that each player attempts to shorten the queue associated with it. For this purpose, they employed Nash equilibrium and Stackelberg equilibrium. Simulations showed that this method improved the length of the queue behind traffic lights by up to 26.45% in comparison to the responsive control[15]. In 2011, Qiao et al. propounded a two-stage fuzzy controller for controlling a signalized intersection. In this research, two objectives of enhancing productivity and achieving equal chance of passage were considered. The controller in the first stage chose the green segment related to the next phase, and in the second stage, it determined its time value. Then, the obtained parameters were optimized using an off-line genetic algorithm. The simulations showed better results compared to previous algorithms[16]. In 2015, Castan et al. presented an agent-based method using propagation neural networks (PNN). This method determined the green range of the traffic light according to the level of demand of the relevant intersection. The experiments on two intersections showed that the extent of mobility of machines increased by up to 28%[17]. In the same year, Araghi et al. first used Cuckoo advanced search algorithm in order to optimize the parameters of traffic lights. In this way, they optimized the smart controller parameters of neural networks (NN), adaptive neuro-fuzzy inference system (ANFIS), and Q-learning (QL), and then compared the results of these controllers with the fixed time controller. NN, ANFIS, and QL functioned 44%, 39%, and 35% better than the fixed time controllers, respectively[18]. In 2015, Long et al. presented a multipurpose optimization model to control the urban traffic network. The variables of the length of the queue behind traffic light and the stop time were considered as the cost function variables. The coefficient of these variables was specified according to an analytical fuzzy process. Next, the obtained function was solved by QL algorithm. Using this method, the program of controlling the intersection lights was generated real-time according to the strategy of interest to planners. Simulations indicated the high efficiency of this method[19]. Subsequently, this method was developed by Clempner and Pozniak based on Markov chain Rule. The effectiveness of this method was proven for different states by stimulating one intersection. The intersection was considered as two players in a non-cooperative game, where each player attempted to reduce its queue. In this way, attempts were made to find the Nash equilibrium[20]. In 2017, Gao et al. used deep reinforcement learning to extract useful data from raw traffic information online and optimal strategy learning for responsive control of traffic lights. They employed experimental responses as well as objective clustering to stabilize the method. The simulations showed better performance of this method (by 46 and 32%) compared to the fixed time and longest queue first method[21]. In 2019, Lu et al. employed explicit model predictive control for controlling signalized intersections. In this method, they first generated some rules for scheduling traffic lights using a multi-parametric quadratic program. Then, according to the traffic data and the prediction performed according to the rule associated with that, the lights were controlled. The results showed the better performance of this method by up to 40% compared to the fixed time method[22].

Overall, the research conducted so far has major problems in three different areas:
1. A large number of studies have founded their modeling based on Webster and Miller, in whose method, it is not possible to model the many states and phenomena associated with the theory of probabilities, while concerning the software developments, this possibility now exists to a large extent. In other words, concerning the available software facilities, there is no need to model changes in the density of streets and variations in turn right and turn left in the form of a specific relation.
2. The current traffic control systems have a control center. More specifically, since local control systems demand large costs, the current systems manage the intersections of traffic network as a central or distributed hierarchical control systems. Therefore, it is not possible to send data in these systems with a high rate (on second scales or even minute scales). Many studies performed so far have been done based on high data submission rate, which cannot be implemented given the infrastructures of the traffic control center.
3. Some of these studies have used novel optimization methods in order to determine the optimal timing of phases. Note that with incorrect mapping of elements associated with the intersection onto the templates that exist in smart methods from the beginning, they have only made the problem and its solution more complex. In other words, the mentioned methods are practical for special conditions and specific agents, which are not comply with an isolated signalized intersection context.

In this research, first different traffic conditions have been modeled based on computational software. According to the modeling performed, 625 different traffic states have been modeled based on the flow rate of the streets leading to the intersection, and the optimal green time for each phase has been specified for each state. Accordingly, the base of rules associated with fuzzy control has been developed.

In the second stage, according to the network data submission rate, the flow rate of each street associated with every piece of data up to the subsequent data has been predicted, based on which the optimal phase magnitude is determined. For example, in the SCATS (Sydney Coordinated Adaptive Traffic System) which is currently used in many countries including Iran for managing the traffic network, it has a data submission rate of 15 min.

In order to determine the optimal phase, two parameters are important: i) the accuracy in phase determination, ii) the speed of achieving the specific value. Note that fuzzy control method has a high speed and low accuracy, while the real-time control method has a low speed and high accuracy. Therefore, in this research, after traffic prediction in the second stage in order to achieve high speed and accuracy in determining the optimal green range, fuzzy and real-time control methods have been combined with each other.

Therefore, briefly in this research considering the network infrastructures and modeling the traffic based on computational software, a method was designed by combining the fuzzy and real-time control, which in comparison with other methods offer the optimal timing with a high accuracy and reasonable duration.

Section 2 presents the issues related to the modeling. In this section, first explanations related to modeling the traffic data have been provided, after which the real-time control method has been described (1-2). Next, the explanations related to the knowledge base and fuzzy control rules' base have been presented (2-2). Eventually, the method of combining these two controllers with each other has been elaborated. Section 3 offers the scenario associated with the streets leading to Abshar intersection in Isfahan and the performance of different controlling methods given the scenarios has been evaluated. Eventually, the results are analyzed and some suggestions are proposed for further research (4).

## 2. Modeling

In this research, with the aim of assessing the algorithms proposed for controlling signalized intersections, a suitable simulation environment has been designed and implemented. Although the current software such as GETRAM (Generic Environment for Traffic Analysis and Modeling), SimTraffic, and Aimsun has a considerable power in simulation, they cannot be used for the case under study here. Instead, the need to a flexible environment for implementing the control algorithms prompted developing a simulation software in a computational environment. In order to validate this model, comparisons were made with AimSun software. In this paper, the mentioned software is called SIFRC.

The main difference between different controlling algorithms lies in applying different logics to determine the decision variables such as the value of phase. Accordingly, in SIFRC, conditions in line with the actual potentials of typical sensors available in traffic control systems have been used. In the environment of this simulator, the vehicles have been shown only based on their existence microscopically, and details such as the type of vehicle, color, and length have been neglected. The reason is that presumably the sensors determine only the existence of a vehicle and its speed at the time of passage.

The number of vehicles approaching the intersection in the green phase and the number of vehicles awaiting in the red phase are generally used in most controlling algorithms. The sensitivity of control algorithms to the values of these variables is a function of the logic governing the control. Therefore, the input variables of SIFRC system include: the number of vehicles entering the intersection in each of the streets leading to it within the data submission time range (based on the current control systems, this range is equal to 15 min or 900 s), the number of vehicles leaving each of the streets leading to the intersection within the previous time range. The output variable of SIFRC system is the green range associated with each phase in the cycle.

The sensors that count the number of vehicles entering the intersection are placed 150 m away from the intersection stop line. The mentioned distance is chosen based on the distance of intersections connected to the intersection, average speed, and capacity of streets, which has been considered as 150 m in this paper. This value in the study by Niatimaki and Persola[23] was considered as 100 m. Also, the number of vehicles leaving the intersection is obtained through the sensors installed immediately prior to the stop line. If the queue length exceeds 150 m (the distance between the first and second sensors), the relevant entrance is set in critical state.

Fig. 1 demonstrates the entrance and exit of vehicles at an intersection as well as its associated sensors alongside the relevant signs. The signs have been completely explained in Table 1.

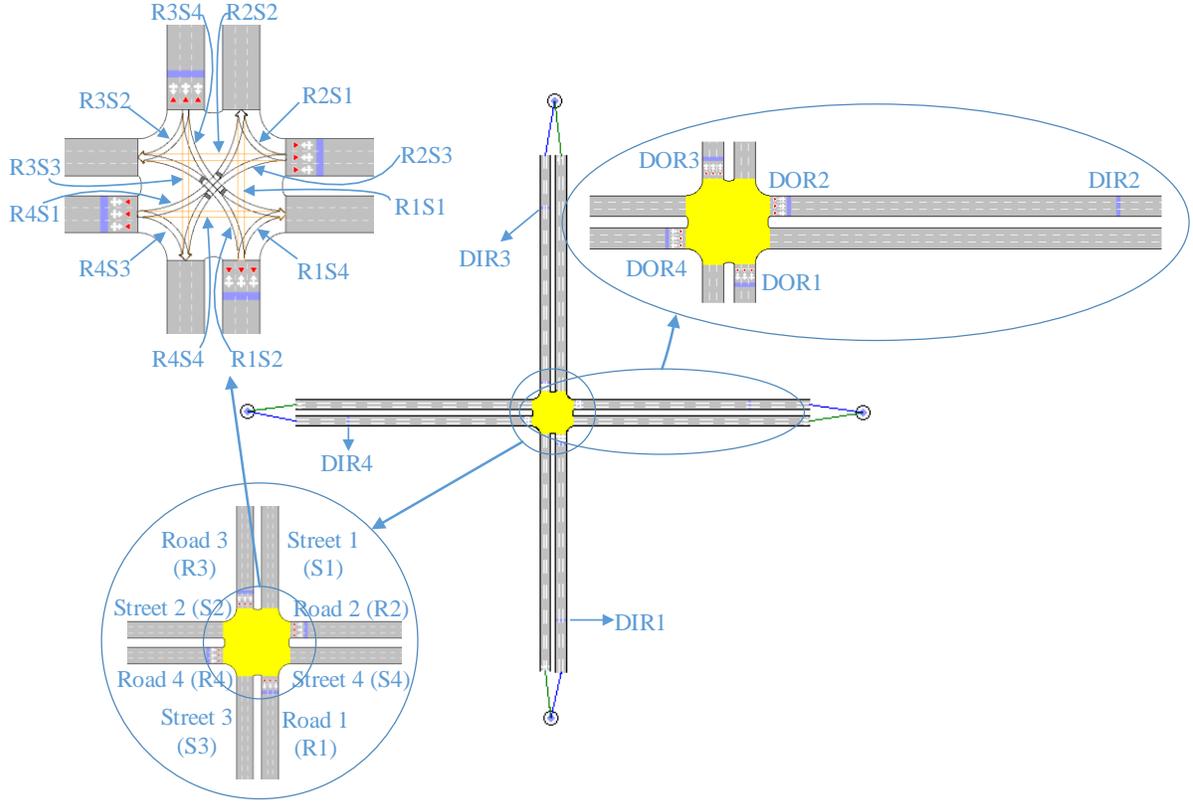

**Fig. 1.** The details related to the signalized intersection

**Table 1.** The signs related to the components of the signalized intersection

| Num | Symbol | Name |
|---|---|---|
| 1 | $CRi\ (i = 1,2,3,4)$ | The capacity of street *i* in every data submission period |
| 2 | $DIRi\ (i = 1,2,3,4)$ | The sensor of vehicles entering the street *i* |
| 3 | $DORi\ (i = 1,2,3,4)$ | The sensor of vehicles leaving the street *i* |
| 4 | $FIRi\ (i = 1,2,3,4)$ | The number of vehicles entering the street *i* in each period of data submission |
| 5 | $FORi\ (i = 1,2,3,4)$ | The number of vehicles leaving the street *i* in each period of data submission |
| 6 | $QRi\ (i = 1,2,3,4)$ | The number of vehicles awaiting in the queue of street *i* |
| 7 | $SQS$ | The sum of vehicles awaiting in the queue of intersection streets |
| 8 | $TRi\ (i = 1,2,3,4)$ | Delay time related to street *i* |
| 9 | $DeRi\ (i = 1,2,3,4)$ | Density of vehicles in street *i* |
| 10 | $T$ | Time unit or time period |
| 11 | $k$ | *k-th* sample |
| 12 | $Ri\ (i = 1,2,3,4)$ | The *i-th* path entering the intersection |
| 13 | $Si\ (i = 1,2,3,4)$ | The *i-th* path leaving the intersection |
| 14 | $RiSj\ \begin{matrix}(i = 1,2,3,4)\\(j = 1,2,3,4)\end{matrix}$ | Percentage of vehicles leaving the *i-th* street and entering the *j-th* path |

On the other hand, in order to generate random numbers in the entrance of vehicles, their distribution should be specified, where random numbers would be made using random variable generation methods based on the distributions. Since the sensors were installed 150 m away from the intersection and the queue line in normal state was shorter than 150 m, in SIFRC software, normal distribution has been used for generation of random numbers. The different traffic rates in the streets leading to the intersection are shown in the Fig. 2.

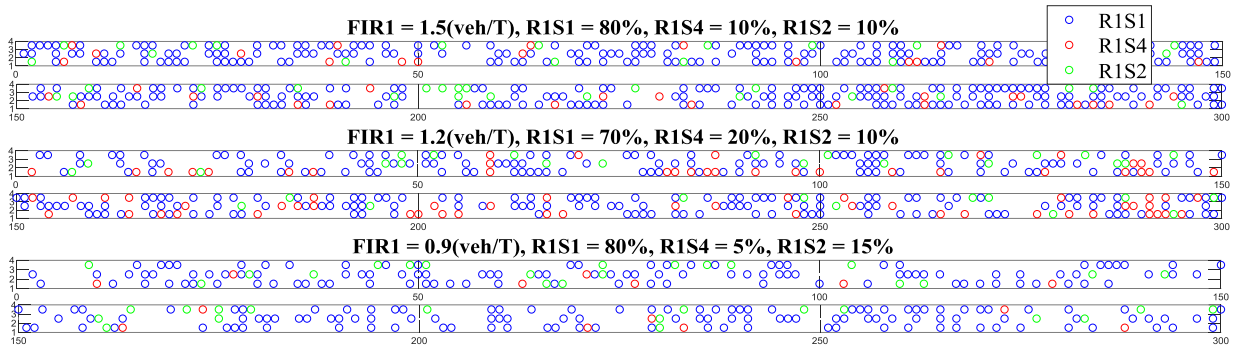

**Fig. 2.** The rate of different flow of vehicles passing through the street along while also segregating the vehicles intending to turn right or left

At this stage, it should be answered whether changes in the distribution of vehicles entering the intersection within the network functioning time ranges make significant changes in the optimal green range. In order to answer this question, 100 repetitions were done for each state and the results were compared with each other. The results showed that these changes in most cases (more than 95%) alter the green range by at most 2 s. On the other hand, the weighted average of data is almost equal to the data with the maximum frequency. In this software, the weighted average of data has been chosen as the optimal value of the green range.

To model the performance of signalized intersections, because of the complexity of intersections and the procedure, the following assumptions have been considered:

1. Every intersection has four incoming and four outgoing streets.
2. The streets have the same size and each street has three identical lines.
3. The maximum capacity for passage of vehicle from each line is equal to one vehicle per unit of time.
4. The incoming and outgoing coefficients are equal to each other, which is 0.5 unit of time per second.
5. The number of vehicles that intend to turn right at each line is equal to half of its right-side line.
6. The number of vehicles that intend to turn left at the intersection in each line is equal to half of its left-side line.
7. In coloring the simulator lights, the total sum of the complete stop, caution time (yellow),

initial waste at the time of initiation of vehicle movements in the queue have been considered as the yellow light time.

Regarding the general course of the modeling, a subprogram is developed to control the signals of intersection with a major body for modeling the entrance and exit of vehicles based on the timing of outflow associated with the signals in the controlling algorithm. The main body of the program involves the following stages:

1. First, the times of entrance of vehicles are generated randomly.
2. For each vehicle, a code is considered which specifies the street and line it is located in at the time of entrance, and when exiting, whether it goes straight ahead or intends to turn light or left.
3. Next, the program related to the light control is implemented and the green time of each phase is specified.
4. After determining the green time of each phase, the red time related to that is also specified in the cycle.
5. Considering the green and red times of each phase, the time of each vehicle exiting data section or its waiting in the waiting queue is determined.
6. Considering the entrance and exit time of each vehicle, the time of the total delay developed by the intersection is calculated.
7. This process is repeated for all of the controlling algorithms.

Based on the above explanations, an intersection has been controlled as a sample as fixed time, and the entrance and exit of vehicles to and from it have been represented for 300 time units. Note that in the mentioned intersection, the initial conditions have been considered as zero. The cycle magnitude is 120 s, yellow time is 4 s, and the intersection has been controlled as two-phase. The first green range is related to streets 1 and 3, while the second green range is associated with streets 2 and 4. The vehicles that exist in street 1 enter the intersection, then wait in the queue, and finally leave it for 300 time units, according to the following figure. It is assumed that vehicle users obey the law when turning right or left off the street. This situation differs given the driving culture of any country, and its value can be adjusted based on different conditions in SIFRC software. Therefore, according to Figs. 3 and Figs. 4, FOR1 and FOR2 parts, the vehicles that intend to turn left should wait until their left side becomes free of vehicle, or their left side vehicle should also intend to turn left. The vehicles that intend to turn right, since according to the law turning right has the right of way for turning left, if the left side vehicle intends to turn right, it should wait until this vehicle turns right; and if there is no vehicle after that or does not intend to turn left, this vehicle is allowed to leave the intersection. Note that the vehicles that leave the intersection directly have the right of way over other vehicles.

According to the Fig. 3, FOR1 diagram, the green phase begins from streets 1 and 3, since there is no queue in its first part, and vehicles have not been waiting in the queue to leave the intersection.

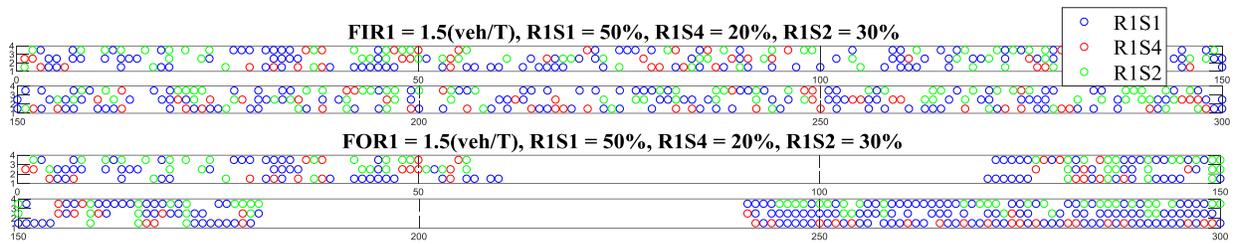

**Fig. 3.** The stages of entrance and exit of vehicles in street 1 (R1) connected to the intersection

The vehicles present in the street 2 and 4 enter the intersection, wait in the queue, and eventually leave it. Fig. 4 shows the arrival and departure of vehicles in the street 2 for 300 time units.

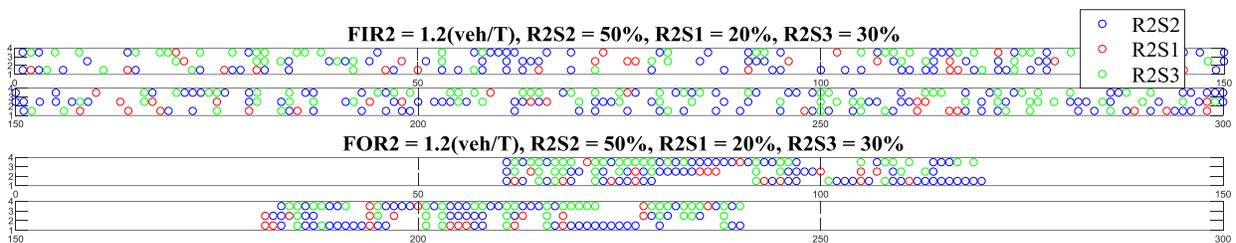

**Fig. 4.** The stages of entrance and exit of vehicles in street 2 (R2) connected to the intersection

The length of the queue developed in streets 1 and 2 per the inputs and outputs represented in the previous figures has been shown in the following figure. As can be observed in the figure, when the traffic light associated with the street becomes red, the queue is elongated, while when the light becomes green, its length decreases.

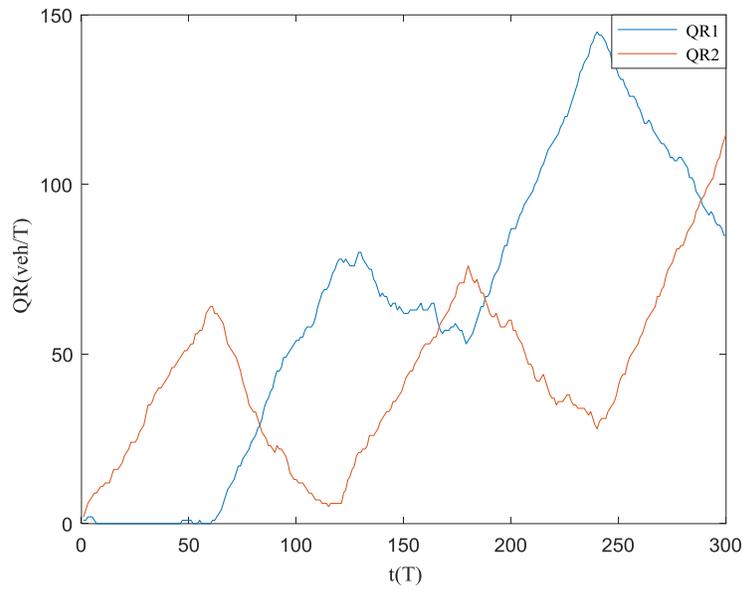

**Fig. 5.** The length of the queue related to streets 1 and 2 in 300 time units

The delay associated with each street is obtained through calculating the area under curve related to its queues' length. This diagram has been shown in Fig. 6.

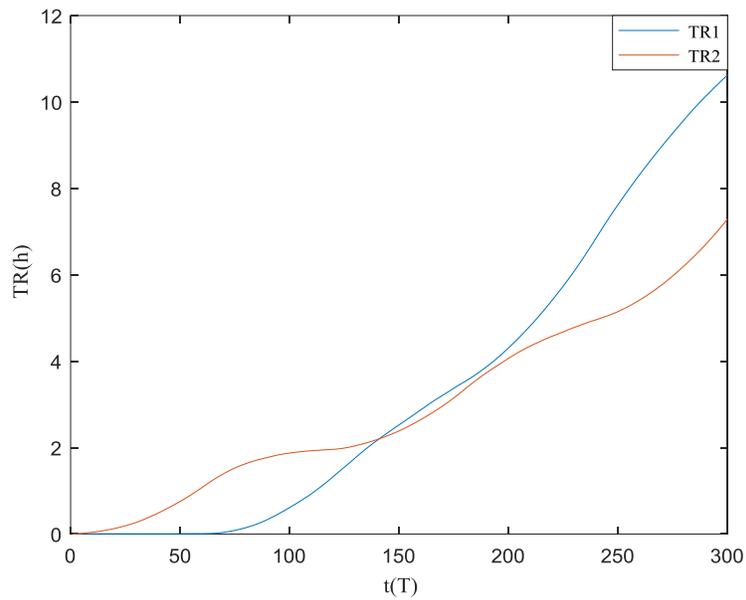

**Fig. 6.** The delay related to streets 1 and 2 in 300 time units

## 3. Real-time control

Using the sensors available in each street, the incoming and outgoing commuting rate in each street is obtained. Using the obtained commuting rates, according to Eq. 1, the value of the vehicles moving between two sensors at the $k_{th}$ time step is obtained.

$$\begin{matrix} DIRi \to FIRi \\ DORi \to FORi \end{matrix} \Rightarrow QRi(k_0) = \sum_{j=1}^{k} FIRi(j) - \sum_{j=1}^{k} FORi(j) \qquad (1)$$

Note that the data submission time unit is every 15 min, while the length of the cycle of traffic lights related to the intersection is equal to 120 s. Thus, there are 7.5 cycles per every 15 minutes; the traffic situation is predicted according to the incoming commuting rate, initial value, as well as the green and red range of each phase in these 7.5 cycles. In other words, QRi($k_0$) is considered as the input related to the first cycle of the $k_{th}$ range. At every time unit of data submission, this input should be reloaded, because the errors related to modeling the output of vehicles are cumulative. More specifically, the sensors installed at the exit part of each street are in charge of readjusting the initial values in each data submission time units.

According to Eq. 2, the density of each street is obtained in every cycle related to the $k_{th}$ time units.

$$DeRi(k_n) = \frac{7.5 QRi(k_n) + FIRi(k)}{CRi} \qquad (2)$$

According to the density obtained in each cycle, the incoming vehicles are modeled based on the explanations given.

Since the length of the cycle has been assumed as 120 s and the number of phases as 2, the delay is calculated for 120 states related to phases in each time range. More specifically, the length of phase 1 ranges between 1 s and 120 s, in line with which the length of phase 2 is curtailed. Based on each state, the outgoing vehicles are modeled, after which the delay developed in the intersection is calculated according to each timing. Definitely, the best timing involves the minimum delay.

Since all of the possible states are tested, the major advantage of this method is achieving the best response. However, the big disadvantage is the duration of achieving this best response. In other words, considering the necessity of modeling for all states, the time of finding the best response becomes relatively long. It can be stated that the best response for resolving the problem related to the response time is fuzzy control.

## 4. Fuzzy control

The fuzzy system is a nonlinear mapper of the vector of input values to a scaler output. The major components of the system include a fuzzifier, base of rules, and defuzzifier[24]. There are a set of 'if-then' rules in the base of rules. After data collection, the inputs should be fuzzified to be usable in the rules base. The fuzzifier part exists in the fuzzy system in order to fulfill this need. On the other hand, the output variable from the 'then' part in the 'if-then' rule, in the base of rules is a fuzzy value, where the outputs of different rules are not necessarily the same. Therefore, for inference, there is a need to a fuzzy inference engine, and to determine the decision variable, the fuzzy output should be converted to a non-fuzzy value. For this purpose, defuzzifier tools are used. Fig. 8 demonstrates the fuzzy control diagram.

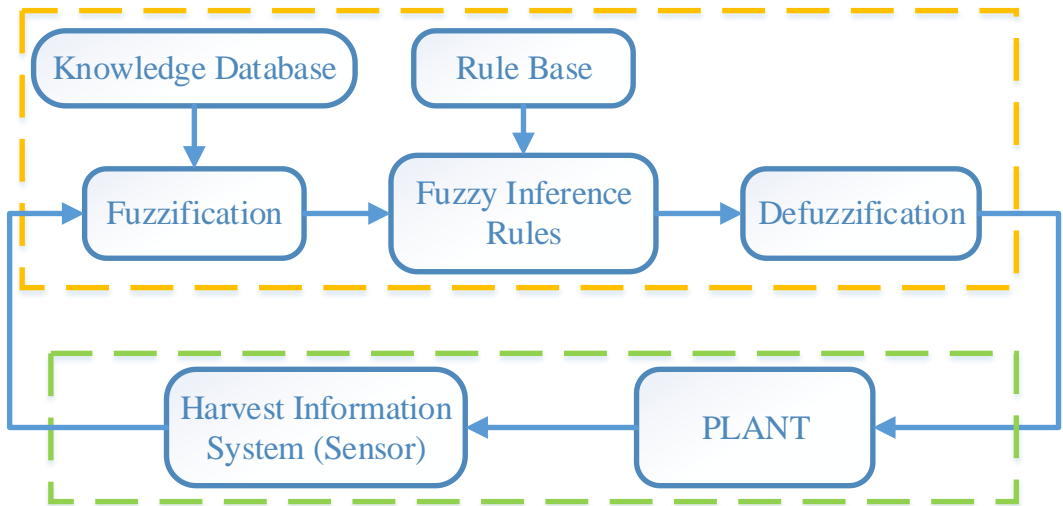

**Fig. 7.** The function of the fuzzy system over signalized intersections

### 2-1. The method of developing membership functions and rules' base

In order to develop membership functions associated with FIRi and QRi variables, which indeed represent the conditions of methods in terms of the number of vehicles approaching the intersection and the vehicles waiting in the queue, first their expected values are calculated. For this purpose, first using the essential Relation 2, the expected density is estimated at an assumed speed and known transit rate. This relation offers an approximate value of the expected density. Next, according to the calculated density and the knowledge base, the corresponding fuzzy values are determined. Assuming normalization of density values associated with each line of the streets leading to the intersection, the maximum density is equal to 1. In the first stage, the density value has been divided into five equal parts ranging from 0 to 1. Assuming that each street has three lines, the fuzzy values related to that are equal to 0.03, 0.9, 1.5, 2.1, and 2.7. These values correspond to the densities of [0.0-0.6], (0.6-1.2, (1.2-1.8), (1.8-2.4] and [2.4-3.0]. Given establishment of equilibrium between the accuracy and volume of the base of rules, the commuting rate of each street was divided into five levels. In this way, since the intersection has four streets and five states have been considered for each street, the intersection finds 625 different phases. If the magnitude of leveling increases considerably, the accuracy of the obtained results also grows. Nevertheless, the number of states is added with the order of 4. As at the end, this method is combined with an real-time control method, usage of knowledge base with further leveling is inessential.

The base of rules has been developed based on SIFRC modeling. Table 1 as an example indicates 10 'if-then' rules of the set of rules in it. For example, the third rule of this table states that: if the density of vehicles in the streets leading to the intersection is 0.3, 2.1, 0.9, and 1.5, then the green range is equal to 3.6.

**Table 2.** Ten samples of the rules related to the fuzzy system rule base

| Num | Road 1 | Road 2 | Road 3 | Road 4 | Green Time |
|---|---|---|---|---|---|
| 1 | 1.5 | 2.7 | 0.3 | 2.7 | 12 |
| 2 | 0.3 | 0.3 | 0.3 | 2.1 | 17 |
| 3 | 0.3 | 1.5 | 0.3 | 0.9 | 19 |
| 4 | 0.3 | 2.1 | 0.9 | 1.5 | 36 |
| 5 | 0.9 | 0.9 | 0.9 | 0.9 | 60 |
| 6 | 0.9 | 2.1 | 1.5 | 2.7 | 36 |
| 7 | 1.5 | 0.9 | 0.3 | 0.9 | 70 |
| 8 | 2.1 | 0.9 | 0.3 | 0.9 | 84 |
| 9 | 2.1 | 0.3 | 0.9 | 0.3 | 103 |
| 10 | 2.7 | 0.3 | 1.5 | 0.3 | 108 |

The major advantage of this method is achieving an optimal response within the shortest possible time, while its main downside is accuracy. In other words, since the density related to each street is mapped over the densities available in the knowledge base, at the end the obtained response is slightly different from the optimal response.

## 5. Fuzzy- real control

Considering the advantages and disadvantages of fuzzy and real-time control methods, it can be concluded that if these two methods are combined properly with each other, their advantages would be kept while their disadvantages would be eliminated. In order to combine these two methods, in this paper considering the density of streets, first using the knowledge base related to the fuzzy control system, the initial timing values have been obtained. Then, considering the real-time control method, the modeling has been performed for 5 seconds shorter and longer than the initial value. In other words, instead of 120 different states, only the delay related to 11 different states has been calculated, and eventually the best response is specified for the timing of traffic lights. Therefore, the real-fuzzy method is 11 times as fast as the real-time method.

## 6. Simulation

In order to investigate the performance of the proposed method, a scenario has been defined for one of the streets leading to a hypothetical intersection from 06:00 a.m. to 22.00. The transit rate of the vehicles for each street in every 15 minutes based on the daily hours and the street being minor or major has been chosen such that it practically creates different states in order to evaluate the performance of different controlling algorithms. The stated transit rates have been equal to the transit rates related to the Abshar intersection in Isfahan city.

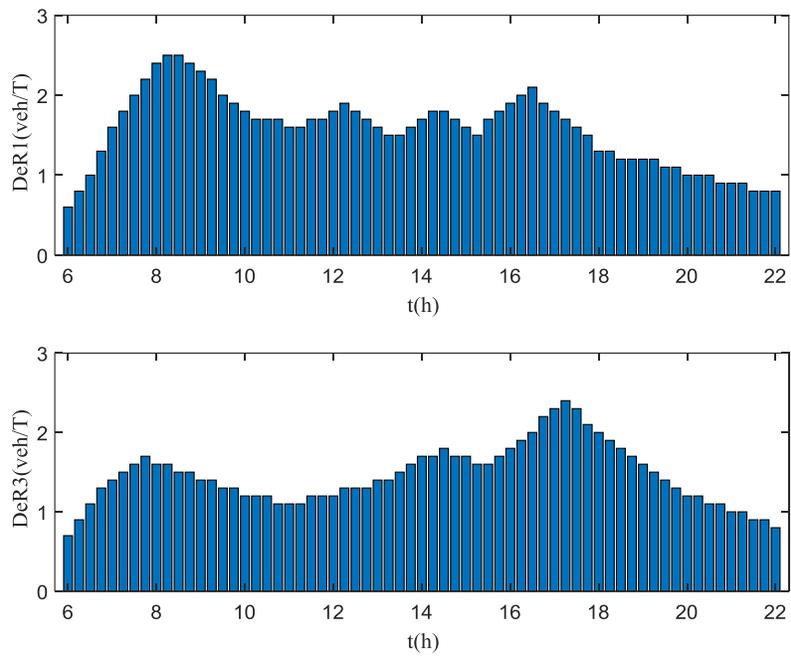

**Fig. 8.** The transit rates related to streets 1 and 3 from 6.00 to 22.00

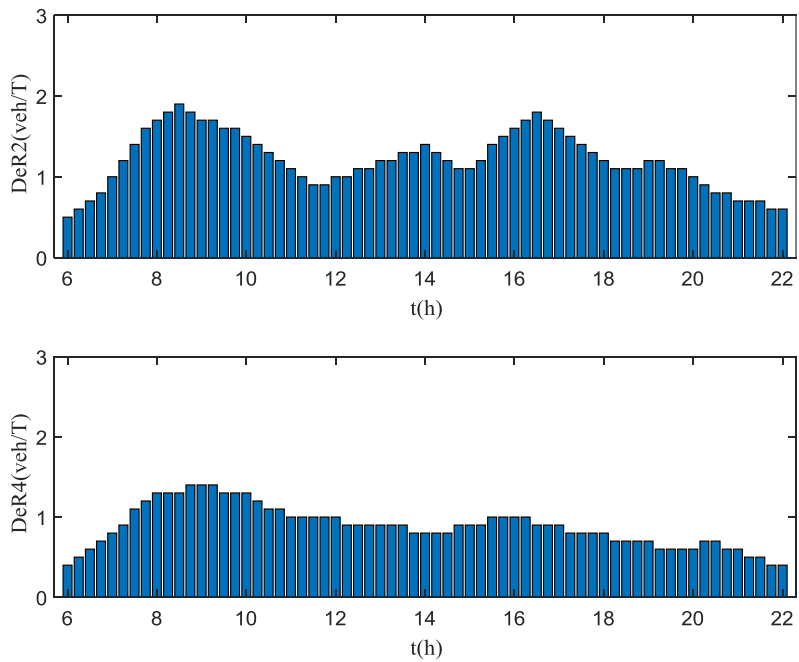

**Fig. 9.** The transit rates related to streets 2 and 4 from 6.00 to 22.00

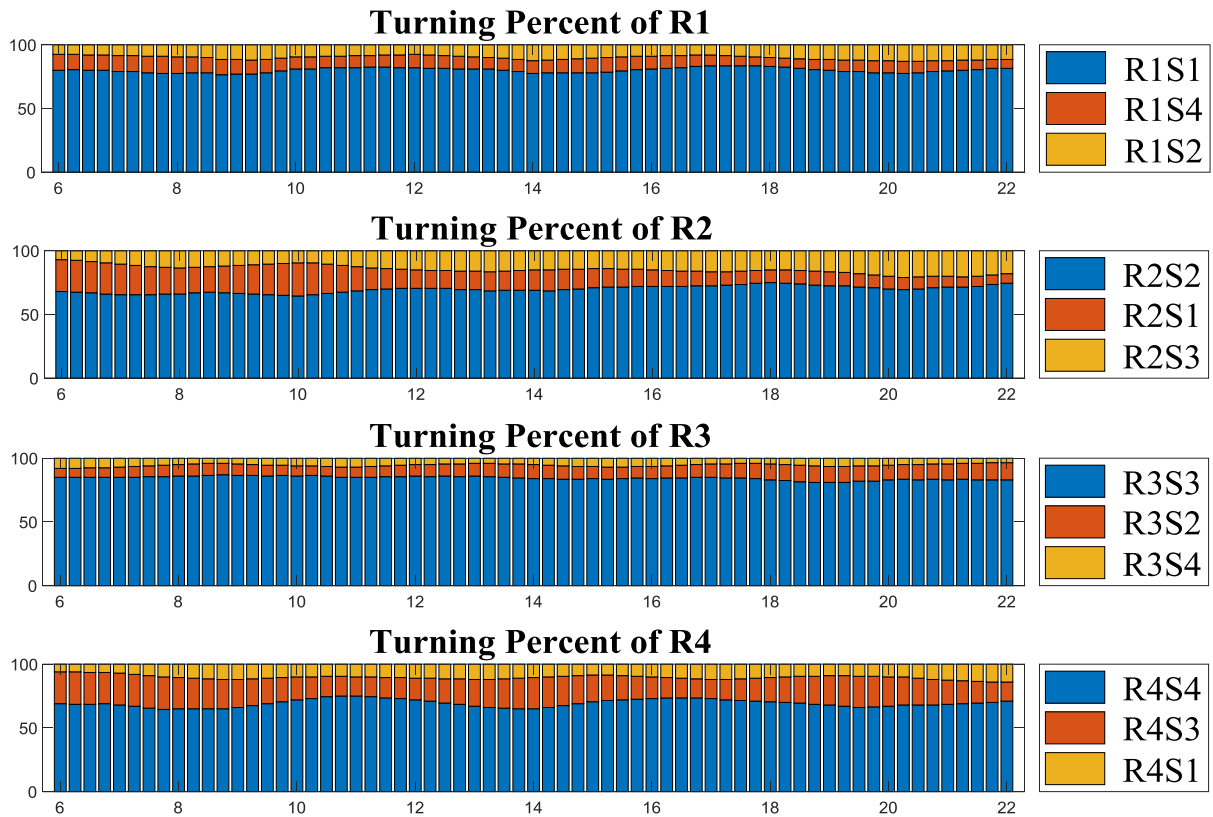

**Fig. 10.** The turn right/left values related to each street

In order to compare the algorithms, identical distributions have been used in order to establish equal conditions in the comparison as the input data. The results have been shown in the following figure for six methods: fixed time, pre-timed, segmental pre-time, fuzzy control, real-time control, and fuzzy- real control.

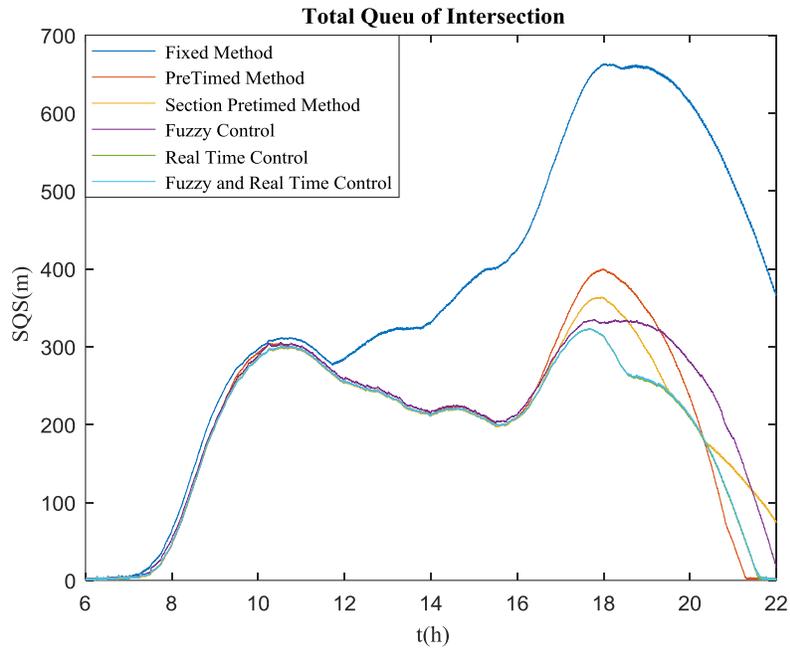

**Fig. 11.** The total length of the queue of the streets connected to the intersection

According to the above figure, the maximum length of the queue has improved by 53% compared to the fixed time, 21% compared to the pre-time, 12% relative to the segmental pre-time, and 6% compared to the fuzzy method.

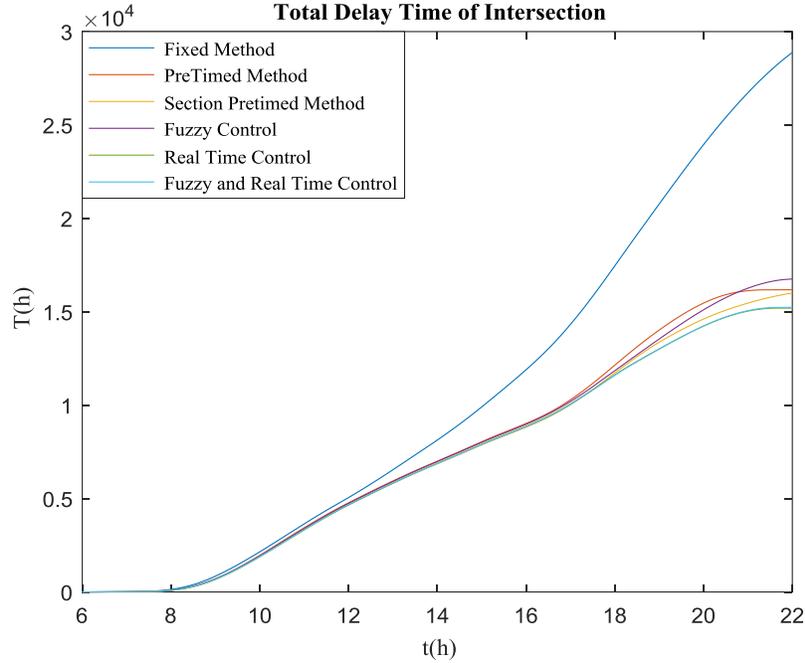

**Fig. 12.** The delay developed by the intersection

According to the above figure, the delay related to the fuzzy- real method has improved by 49% and 7% compared to fixed time and other methods, respectively. As can be seen in Figs. 11 and 12, the diagram related to the real-time control has become similar to the diagram associated with the fuzzy- real control. Therefore, the response of these two methods is the same, while the speed of achieving the optimal response in the fuzzy- real control has grown by 11 times. The value of this method is further highlighted when it is compared with the extent of impact of other methods. According to [18], Cuckoo-neural, ANFIS-Cuckoo, and QL-Cuckoo methods offered improvement by at most 44, 39, and 35% respectively compared to the fixed time method. Based on [17], multi agent PNN algorithm improved the delay time by at most 20% compared to the fixed time method. According to [15], the game theory algorithm reduced the delay by 26.45%. Based on [21], the deep reinforcement learning algorithm decreased the delay time by 46% compared to the fixed time method. As can be deduced from these papers and many other similar papers, in spite of applying complex methods, they have not been as efficient as the method propounded in this paper.

## 7. Conclusions and suggestions

In controlling traffic, several traffic flows compete with each other over time and space, but there are often different criteria for traffic flows. Generally, the criterion of delay is the determinant factor in the efficiency of traffic control systems.

In this paper, Abshar intersection in Isfahan city was used as the experimental sample. The data membership functions were specified, the rule base was established using SIFRC, and then by presenting a new fuzzy- real strategy to determine the optimal phase time, the performance of the intersection was evaluated. The results indicated the better performance of this method by up to 49% compared to fixed time and 7% relative to pre-timed and fuzzy methods.

Since in this paper the performance of the intersection was investigated independently of the network, the number of possible states was limited. Nevertheless, in order to match this intersection with its surrounding intersections, a large number of states should be investigated in order to determine the optimal offset time alongside the optimal phase. In this regard, a possible solution is limiting the number of states in a regulatory way or using optimization algorithms. Therefore, a subject that can be proposed for future research is stating an optimal method to match intersections with each other with the aim of minimizing the delay. Therefore, future research should design a method which can combine the fuzzy- real methods implemented in each intersection with each other using smart algorithms.